\documentclass[aps,prl,twocolumn,superscriptaddress,showpacs]{revtex4}

\usepackage{graphicx}
\usepackage{dcolumn}
\usepackage{bm}
\usepackage{color}

\begin{document}
\preprint{APS/123-QED}

\title{Unusual coexistence of negative/positive charge-transfer in mixed-valence Na$_x$Ca$_{1-x}$Cr$_2$O$_4$}

\author{M. Taguchi}

\email[Corresponding author: ]{mtaguchi@ms.naist.jp}

\affiliation{Material Science, Nara Institute of Science and Technology (NAIST), Ikoma, Nara, 630-0192, Japan}

\author{H. Yamaoka}

\affiliation{RIKEN SPring-8 Center, Sayo, Sayo, Hyogo 679-5148, Japan}

\author{Y. Yamamoto}

\affiliation{School of Science and Technology, Kwansei Gakuin University, Sanda, Hyogo 669-1337, Japan }

\author{H. Sakurai}

\affiliation{ National Institute for Materials Science 1-1 Namiki, Tsukuba 305-0044 Japan}

\author{N. Tsujii}

\affiliation{ International Center for Materials Nanoarchitectonics (MANA), National Institute for Materials Science, 1-2-1 Sengen, Tsukuba, Ibaraki 305-0047, Japan}

\author{M. Sawada}

\affiliation{Hiroshima Synchrotron Radiation Center, Hiroshima University, Hiroshima 739-0046, Japan}

\author{H. Daimon}

\affiliation{Material Science, Nara Institute of Science and Technology (NAIST), Ikoma, Nara 630-0192, Japan}

\author{K. Shimada}

\affiliation{Hiroshima Synchrotron Radiation Center, Hiroshima University, Hiroshima 739-0046, Japan}

\author{J. Mizuki}

\affiliation{School of Science and Technology, Kwansei Gakuin University, Sanda, Hyogo 669-1337, Japan}

\date{\today} 

\begin{abstract}
We have investigated the electronic structure of Na$_x$Ca$_{1-x}$Cr$_2$O$_4$  using x-ray absorption spectroscopy together with Anderson impurity model calculations with full multiplets. We show Na$_x$Ca$_{1-x}$Cr$_2$O$_4$ taking a novel mixed-valence electronic state in which the positive charge-transfer (CT) and the negative self-doped states coexist. While CaCr$_2$O$_4$ (one end member) exhibits a typical CT nature with strong covalent character, Na substitution causes a self-doped state with an oxygen hole. In NaCr$_2$O$_4$ (the other end member), positive CT and negative self-doped states coexist with equal weight. This unusual electronic state is in sharp contrast to the conventional mixed-valence description, in which the ground state can be described by the mixture of Cr$^{3+}$ ($3d^3$) and Cr$^{4+}$ ($3d^2$). 
 
\end{abstract}


\maketitle
Correlated metal oxides show a wide variety of physical properties; for example, high-temperature superconductivity, colossal magneto-resistance (CMR), multi-ferroelectricity, and metal-insulator transitions, among others\cite{ima98}. Such a variety of complex phenomenon are derived from $d$-electron duality; $i.e.$ competition or interplay of the localized and itinerant degrees of freedom of the $d$ electrons. For this reason, most of the exotic properties appear at the border between metallic and insulating states. Thus, the study of electronic states in the insulating states is of crucial importance towards the understanding of the physics of the strongly correlated electron systems. In the context of an electronic state, the insulating states are caused by three factors; namely, the $d$-$d$ Coulomb repulsion energy $U$, the charge-transfer energy $\Delta$, the $p$-$d$ hybridization energy $V$ between oxygen $p$ and transition metal $d$ orbitals. Thus, as Zaanen, Sawatzky, and Allen (ZSA) proposed\cite{zaa85}, there are two different types of charge gaps for the insulating states: Mott-Hubbard type and charge-transfer (CT) type. For the former type, a metallic $d$ band splits into lower and upper Hubbard bands with an insulating gap determined by the size of $U$. By contrast, in some cases, the oxygen $p$-bands are located between the Hubbard bands in the situation that the CT energy $\Delta$ is smaller than the on-site, $U$. In this case, the charge gap is comparable not with $U$, but with $\Delta$. These can be schematically summarized in a ZSA phase diagram\cite{zaa85,boc96} (Fig.~1(a)). It is apparent that the negative CT regime appears in the phase diagram\cite{miz91,saw16}, in which the CT energy is further reduced, and $d^n$ electronic configuration ($n$: the number of the $d$-electrons) is no longer stable. Instead, the $d^{n+1}\underline{L}$ electronic configuration ($\underline{L}$: a hole at oxygen ions) is realized, resulting in a difference in the valence of the transition metal ions from the formal valence. In other words, a finite density of holes, $\underline{L}$, is self-doped into the ligand band and the transition metal takes on a $d^{n+1}$ configuration. Mixed-valence compounds lie in close proximity to a narrow boundary sandwiched between the positive CT regime and the negative self-doped regime\cite{saw16}. Its electronic ground state is usually described as a mixture of $d^n$ and $d^{n+1}$ formal configurations with very small positive CT energy $\Delta$\cite{saw16}. Here we report a novel state in the electronic configurations for a mixed-valence insulator.

\begin{figure}
\includegraphics[scale=.45]{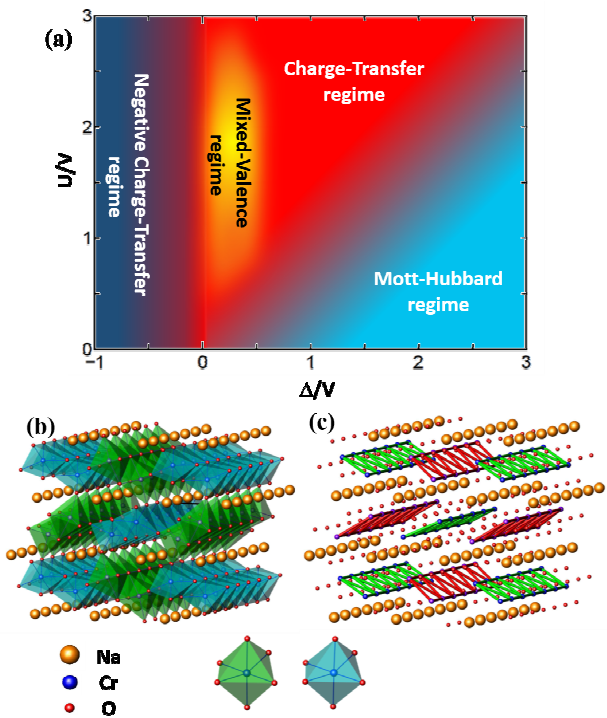}
\caption{\label{fig1}   
(Color online)  Revised Zaanen-Sawatzky-Allen phase diagram and crystal structure of NaCr$_2$O$_4$. (a), Revised Zaanen-Sawatzky-Allen diagram. (b) Crystal structure of NaCr$_2$O$_4$, characterized by two inequivalent Cr$^{3+}$ sites with slightly different oxygen coordination.  Focusing only on Cr sites with magnetic moments, the geometrically frustrated double-chain structure can be seen, as shown in (c).   }
\end{figure}

A Calcium ferrite-type NaCr$_2$O$_4$ (formally NaCr$^{3+}$Cr$^{4+}$O$_4$) is a new member of the chromium oxides with a mixed-valence state of formally Cr$^{3+}$/Cr$^{4+}$ cations. The compound crystallizes in the orthorhombic space group, Pnma, in which Cr cations are coordinated by oxygen octahedrons (see Fig.~1(b)). The double chains of edge-sharing CrO$_6$ octahedrons form zig-zag chains of Cr ions, in which geometrical frustration is induced, and the double chains are connected with each other to form other zig-zag chains of Cr ions between two chains (see Fig. 1(c)). NaCr$_2$O$_4$ is electrically insulating and shows a canted antiferromagnetic transition at T$_N$ = 125 K with spin frustrations in the zig-zag chains. Remarkably, it exhibits an unconventional CMR effect below T$_N$, in which almost -100$\%$ magnetoresistance has been observed by application of 9 T magnetic field at a temperature of 50 K\cite{sak12}. The CMR effect in NaCr$_2$O$_4$ persists down to 2 K and appears within the antiferromagnetic insulating phase, in sharp contrast to conventional CMR in the manganese oxides ($i.e.$, the conventional CMR effect appears only around a ferromagnetic transition temperature). The mixed-valence state and the spin frustrations in the chains are considered as the origin of the unconventional CMR effects\cite{sak12}.	

\begin{figure}
\includegraphics[scale=.45]{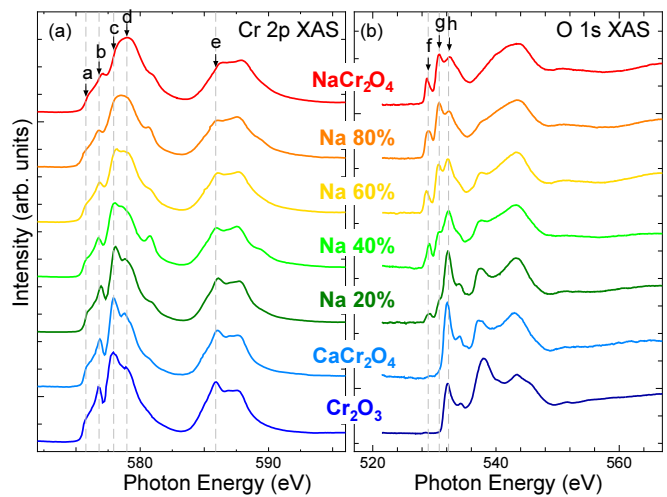}
\caption{\label{fig2}   
(Color online)  Measured evolution of XAS spectra for Na$_x$Ca$_{1-x}$Cr$_2$O$_4$, together with the experimental spectra of the Cr$_2$O$_3$. (a) Na doping dependence of the experimental Cr 2$p$ XAS spectra measured at room temperature. Areas under spectral curves in all spectra were normalized to the same value. (b) Na doping dependence of the experimental O 1$s$ XAS spectra measured at room temperature.  }
\end{figure}

In this study, we show a novel type of mixed-valence state in chromium oxides ($i.e.$ coexistence of positive and negative charge-transfer (CT) states with oxygen hole due to hole doping) by means of high-resolution x-ray absorption spectroscopy (XAS) technique. This state can be expected to play a central role in the unconventional CMR phenomenon of NaCr$_2$O$_4$.

 We employed polycrystalline samples of Ca$_{1-x}$Na$_x$Cr$_2$O$_4$  ($0 \leq x \leq 1$). The details of sample synthesis are reported in Refs. \cite{sak12,sak14} and Supplemental Materials\cite{sup}. XAS measurements were performed on the beamline (BL-14) at Hiroshima Synchrotron Radiation Center (HiSOR), Hiroshima University with the total electron yield mode, where the polycrystalline samples of Na$_x$Ca$_{1-x}$Cr$_2$O$_4$ were fractured under vacuum\cite{saw07}. Figures~2(a)-(b) present an overview of XAS at the Cr $L_{2,3}$-edge and O $K$-edge at room temperature for Na$_x$Ca$_{1-x}$Cr$_2$O$_4$, respectively. Cr $2p$ XAS spectra show complex structures dominated by multiplets, crystal-field and covalence effects, suggesting strong electron correlations. The O $1s$ XAS spectra exhibit a triple-peaked sharp structure between 528 and 535 eV and a broader structure around $535\sim550$ eV. The triple-peaked structures are generally attributed to O $2p$ states strongly hybridizing with unoccupied Cr $3d$ orbitals or hole states in the oxygen sites. Clear and systematic changes in the spectral shape were observed as a function of Na substitution; particularly, in the intensity of the spectral features labeled a$\sim$e in the Cr $2p$ XAS spectra and f$\sim $h in the O $1s$ XAS spectra (see Figs.2(a) and 2(b)).  
It should be noted that the spectral shape of CaCr$_2$O$_4$ bears marked similarity to that of Cr$_2$O$_3$ (prototype material for Cr$^{3+}$), where Cr $3d$ states mix strongly with the oxygen $2p$ state with positive charge transfer energy. The Cr (III) oxides, such as Cr$_2$O$_3$, have a significant charge-transfer insulator character owing to the strong O $2p$ - Cr $3d$ hybridization\cite{uoz97}, although they are sometimes wrongly classified as Mott-Hubbard insulators together with Ti or V oxides. Therefore, the close similarity of CaCr$_2$O$_4$ and Cr$_2$O$_3$ suggests that CaCr$_2$O$_4$ will be a highly covalent charge-transfer system. Our modeling of Cr $2p$ XAS data for CaCr$_2$O$_4$ using the full-multiplet calculations also suggests this highly covalent character with positive charge transfer energy. 

\begin{figure}
\includegraphics[scale=.45]{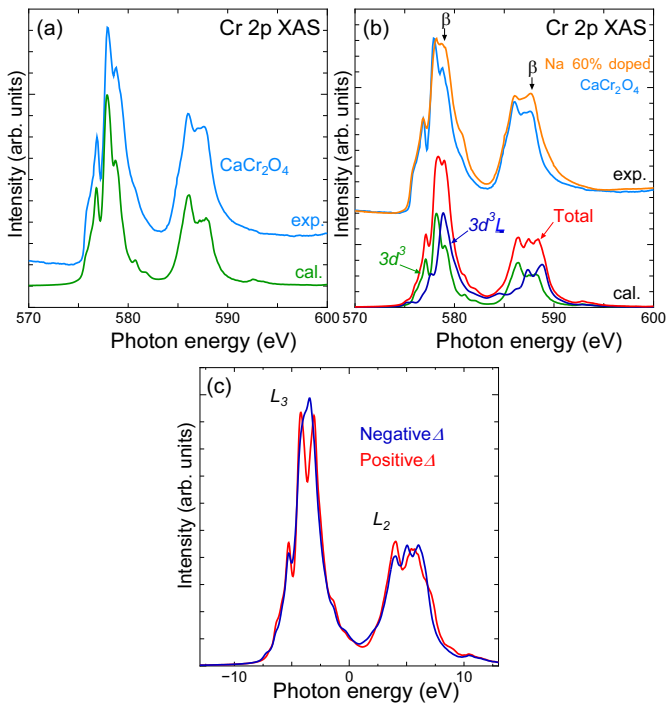}
\caption{\label{fig3}   
(Color online)  Comparison between AIM calculations and experiments for Cr 2$p$ XAS spectra. (a) Comparison of Cr 2$p$ XAS spectrum for CaCr$_2$O$_4$ between calculated (green line) and experimental (light blue line) spectra. (b) To clarify the effect of Na doped-induced changes in Cr 2$p$ XAS spectra, Cr 2$p$ XAS spectrum for Na 60$\%$ doped sample is compared with one for CaCr$_2$O$_4$ (upper panel). The spectra are normalized by maximum intensity. (c) Comparison of calculations between positive and negative $\Delta$, simply by changing the sign of $\Delta$ for hole doped site calculation. The other parameter values are fixed to the values in the supplemental material\cite{sup}.  }
\end{figure}

The Cr $2p$ XAS spectrum for CaCr$_2$O$_4$ is perfectly reproduced by the conventional single Anderson impurity model (AIM)  with full multiplets, as shown in Fig.~3(a). Details of the model calculations and Hamiltonian have been described in Supplemental Material\cite{sup} and previous works\cite{kot01,oga00,cow81,oka95}. By this model, the charge-transfer energy and on-site Coulomb energy were estimated to be $\Delta$$=$$5.8$ eV and $U$$=$$5.5$ eV, respectively. Given that these values are comparable with each other, CaCr$_2$O$_4$ is indeed classified as an intermediate-type insulator between charge-transfer type and Mott-Hubbard type insulator. With the parameter values given above, the ground state is the mixed state of the three configurations with the mixing weights $45.3\%$ ($3d^3$), $44.6\%$ ($3d^4\underline{L}$), and $10.1\%$ ($3d^5\underline{L}^2$), resulting in the $3d$-electron number of $n_d \simeq 3.6$ at the ground state. Here $\underline{L}$ represents a hole in the oxygen octahedron sites. These values clearly indicate that the electronic state of the ground state is a strong mixture of $3d^3$ and the charge transferred $3d^4\underline{L}$ configurations originating from the strong $p$-$d$ hybridization.

\begin{figure}
\includegraphics[scale=.45]{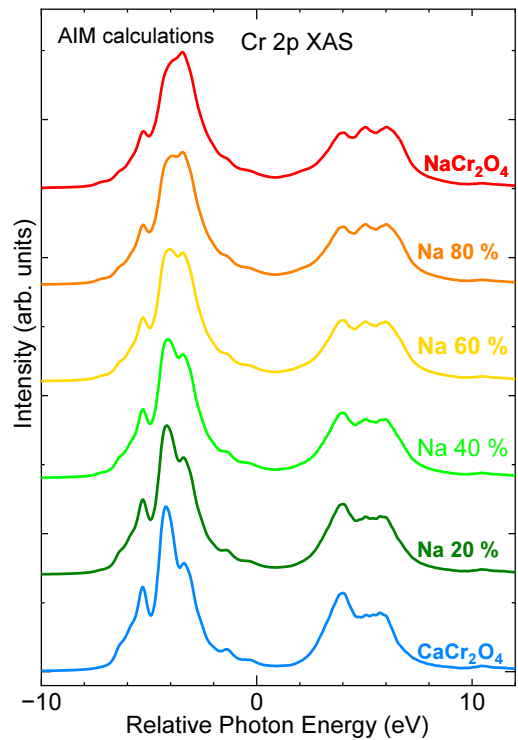}
\caption{\label{fig4}   
(Color online)  Simulated evolution of XAS spectra for Na$_x$Ca$_{1-x}$Cr$_2$O$_4$ as a function of Na doping level. The doping dependence of the total spectrum was obtained by making a superposition of two calculated spectra in accordance with Na doping level.  }
\end{figure}

We now focus on the Na substitution dependence of the absorption spectra. To clarify the effect of Na substitution, the measured absorption spectra of CaCr$_2$O$_4$ and Na $60\%$ doped CaCr$_2$O$_4$ are compared in the upper panel of Fig.~3(b). The spectral shapes for Na$_{0.6}$Ca$_{0.4}$Cr$_2$O$_4$ are clearly different than those of CaCr$_2$O$_4$. Especially, the spectral weights of the higher photon energy peaks labeled $\beta$ in the main line of $2p_{3/2}$ and $2p_{1/2}$ edge are strongly enhanced by Na substitution. As shown in Fig.~3(b), these new features cannot be reproduced by the AIM calculation based on three ($3d^3$, $3d^4\underline{L}$, and $3d^5\underline{L}^2$) basis configurations. The difference matches rather well with the calculated spectrum for hole doped formally tetravalent Cr$^{4+}$ states described by a linear combination of $3d^2$, $3d^3\underline{L}$, and $3d^4\underline{L}^2$ basis configurations, shown as a blue line in the lower panel of Fig.~3(b). Since the Na substitution for Ca introduces holes in the electronic state, a major difference between CaCr$_2$O$_4$ and the Na-doped CaCr$_2$O$_4$ arises from the partial oxidation of Cr$^{3+}$ to Cr$^{4+}$. The ground state of hole doped sites are the mixed state of the three configurations with the mixing weights $8.3\%$ ($3d^2$), $57.6\%$ ($3d^3\underline{L}$), and $34.1\%$ ($3d^4\underline{L}^2$), resulting in the ground state $3d$-electron count $n_d \simeq 3.3$. Thus, to calculate the spectrum for Ca$_{0.4}$Na$_{0.6}$Cr$_2$O$_4$, the spectra for the trivalent and tetravalent ions needs to be superposed with a relative weight of $40\%$ and $60\%$, respectively. This calculation is reliable because the spectra for all the compounds with different Na contents were reproduced just by adjusting the relative weight according to the Na contents, as shown in Fig.~4. It should be emphasized that the agreement between the calculation and observation is not only on the intensity but also on the peak positions.

It is important to realize that the charge transfer energy $\Delta$ for hole-doped Cr sites becomes negative, because $\Delta$ for the tetravalent state is reduced from that for the trivalent state by approximately the same energy scale with Coulomb repulsion, $U$. As a result, the average energy for the charge transfer state $E(3d^3\underline{L})$ is lower than that for $E(3d^2)$, suggesting a strong mixture of $3d^2$ and $3d^3\underline{L}$ configurations with a large weight of negative charge transfer $3d^3\underline{L}$ in the ground state. Because the $\underline{L}$ represents a hole in the oxygen $2p$ orbital in the CrO$_6$ octahedron, a generated hole by Na substitution is primarily trapped in the oxygen sites rather than in Cr sites, just as in NaCuO$_2$\cite{miz91}, NdNiO$_3$\cite{miz00,joh14,bis16}, and NiGa$_2$S$_4$\cite{tak07}. As a counter check, we also made a calculation for NaCr$_2$O$_4$ by a superposition of the trivalent and tetravalent sites with positive $\Delta$ spectra, which corresponds to the conventional mixed-valence picture ($i.e.$ mixture of $3d^2$ and $3d^3$). The calculation clearly disagrees with the observation, as seen in Fig.~3(c), in particular in the $2p_{3/2}$ edge energy region. This distinctly indicates that the conventional mixed-valence picture is not applicable.

\begin{figure}
\includegraphics[scale=.43]{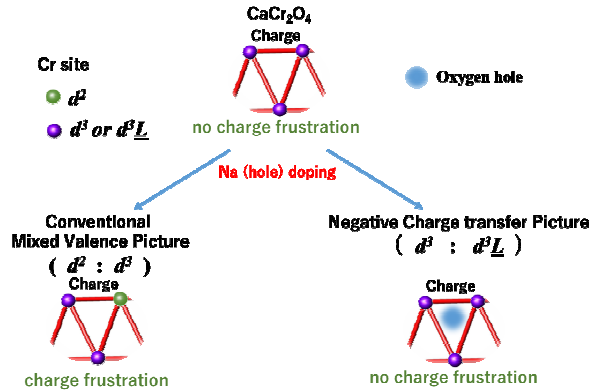}
\caption{\label{fig1}   
(Color online)  Illustration of hole induced behavior of charge frustration. The $3d^3$ and $3d^3\underline{L}$ states are represented by the same purple balls. Because Cr 3$d$ has the same electron number in either state. Hole doping in the Cr site leads to the formation of charge frustration or charge ordering. On the other hands, there is no charge frustration at Cr sites by doping the hole to the oxygen octahedron. }
\end{figure}

Additionally, if the formally tetravalent Cr ions would be in the $d^2$ electronic configuration, several charge ordering patterns have the same energy regarding Coulomb repulsion among Cr ions, which causes charge frustrations (see Fig.~5). The ligand holes may be stabilized to relax the charge frustrations by being shared with some Cr ions, as shown in Fig.~5.  Sharing of the ligand holes by Cr ions has been confirmed by $^{53}$Cr nuclear magnetic resonance (NMR) measurement\cite{tak13}. We also note that the electronic state of the correlated metal oxide usually has either positive or negative $\Delta$ states. Therefore, it is surprising that positive and negative $\Delta$ states coexist in the single compound of NaCr$_2$O$_4$.

One of the most intriguing aspects of this highly unusual self-doped state is its electronic properties. Notably, the $3d$-electron counts, $n_d$, at hole-doped sites remain the same as the ones at trivalent sites even though the holes are introduced. This fact is consistent with the $^{53}$Cr NMR measurement\cite{tak13}, as mentioned above. Furthermore, the coexistence of $3d^3$ with positive $\Delta$ and $3d^3\underline{L}$ with negative $\Delta$ (oxygen hole states) may be responsible for a crossover from $p$ to $n$-type charge carriers at T=230 K in NaCr$_2$O$_4$ which was observed in the temperature evolution of the Seebeck coefficient\cite{kol13}. Subsequently, in reviewing the O $1s$ XAS spectra in Fig.~1b, it is clear that the increase in features $f \sim h$ can be viewed as evidence for negative charge transfer energetics (holes in the oxygen sites), just as in Cuprate and Ni$_x$Li$_{1-x}$O \cite{yar87,nuc88,kui89,che91}. 

In conclusion, the high energy resolution XAS is used to show that the extraordinary coexistence of positive and negative CT states ($i.e.$ unconventional mixed-valence), which leads to unusual CMR in NaCr$_2$O$_4$. As indicated above, the high oxidation state of Cr$^{4+}$ tends to take the negative $3d^3\underline{L}$ state by generating a hole in oxygen octahedron sites. Therefore, the hopping of the ligand holes will considerably contribute to the electrical conduction of this compound. In fact, $p$-type conductivity has been observed by Seebeck measurements below T$_N$, where the CMR effect occurs\cite{kol13}. Thus, the occurrence of the CMR effect indicates that the holes become easier to hop between oxygen ions with increasing magnetic field. Because the holes are magnetically coupled with the antiferromengatically ordered Cr moments, the significant enhancement of the hole hopping strongly suggests that the magnetic structure of the compound is sensitively affected by the application of a magnetic field.

This work was partially supported by KAKENHI (Grants No. 25400335).
The experiments at HiSOR BL-14 were performed under Proposal Nos. 14-A-4 and 14-B-17.

\end{document}